\documentclass[numbook, envcountsect, envcountsame, envcountreset, smallextended]{svjour3}                 
\usepackage{mathptmx,amsmath,amsfonts,stmaryrd,amssymb,textcomp,keyval,latexsym}
\newcommand \R {\mathbb{R}}
\newcommand \N {\mathbb{N}}
\newcommand \D {\mathbb{D}}
\renewcommand \P {\mathbb{P}}
\newcommand \E {\mathbb{E}}
\newcommand \G {\mathrm{Greek}}
\newcommand \C {\mathcal{C}}
\renewcommand \L {\mathbf{L}}
\newcommand \Hh {\mathcal{H}}
\newcommand \lc{\leq_c}

\def\jelclassname{{\bfseries JEL Classification:}\enspace}
\def\jelclass#1{\par\addvspace\medskipamount{\rightskip=0pt plus1cm
\def\and{\ifhmode\unskip\nobreak\fi\ $\cdot$
}\noindent\jelclassname\ignorespaces#1\par}}
\begin{document}
\title{ Smart expansion and fast calibration for jump diffusion}

\author{E. Benhamou         \and
        E. Gobet        \and
        M. Miri
}

\institute{   E. Benhamou, M. Miri \at
              Pricing Partners, 204 rue de Crim\' ee
              75019, Paris, FRANCE.
              \email{eric.benhamou@pricingpartners.com, mohammed.miri@pricingpartners.com}. 
              \and
              E. Gobet, M. Miri \at
              Laboratoire Jean Kuntzmann, Universit\' e de Grenoble and CNRS, BP 53, 38041 Grenoble cedex 9, FRANCE.
              \email{emmanuel.gobet@imag.fr}.
}

\date{First version: May 11, 2007/ This version: September 17, 2008}

\maketitle
\begin{abstract} 
Using Malliavin calculus techniques, we derive an analytical formula for the price of European options, for any model including local volatility and Poisson jump process. We show that the accuracy of the formula depends on the smoothness of the payoff function. Our approach relies on an asymptotic expansion related to small diffusion and small jump frequency/size. Our formula has excellent accuracy (the error on implied Black-Scholes volatilities for call option is smaller than 2 bp for various strikes and maturities). Additionally, model calibration becomes very rapid.
\end{abstract}
 \jelclass{G13}
\section{Introduction}
\label{intro}
The standard Black-Scholes formula (1973) was derived under the assumption of lognormal diffusion with constant volatility to price calls and puts. However, this hypothesis is unrealistic under real market conditions because we need to use different volatilities to equate different option strikes $K$ and maturities $T$. Market data shows that the shape of the implied volatilities takes the form of a smile or a skew.
 
In order to fit the smile or the skew, Dupire (in \cite{Dup94}) and Rubinstein (in \cite{Rub94}) use a local volatility $\sigma_{loc}(t,f)$ depending on time $t$ and state $f$ to fit the market. This hypothesis is interesting for hedging because it maintains the completeness of the market. However, in a few cases \cite{ACGL01}, one has closed formulas. In the case of homogeneous volatility, singular perturbation techniques in \cite{HW99} have been used to obtain asymptotic expression for the price of vanilla options (call, put). Other cases have been derived using an asymptotic expansion of the heat kernel for short maturity (see \cite{Lab05}).

But Andersen and Andreasen in \cite{AA00} show that this sole assumption of local volatility is not compatible with empirical evidence (for instance, the post-crash of implied volatility of the S\&P500 index). Hence, they derived a model with local volatility plus a jump process to fit the smile (we write it AA model). Their model may be seen as a perturbation of pure local volatility models. Of course, this is not the only alternative modelling\footnote{for instance, see the book by Lewis \cite{lewi:00} on stochastic volatility models or the one by Gatheral \cite{Gat06} on models explaining the volatility surface.}. The AA model fits some market data well (see calibration results in \cite{AA00} and those in this work), although we are aware that it does not work systematically nicely. In the following, we do not discuss the relevance of this model in specific situations. We simply focus on this model in order to illustrate our new approach for  numerical pricing and fast calibration. For an analogous study on the time dependent Heston model, we refer to our work in preparation \cite{benh:gobe:miri:08}. Andersen and Andreasen \cite{AA00} calibrate their model by solving the equivalent forward PIDE. This sort of problem could be handled numerically using: an ADI-FFT scheme in \cite{AA00}, a Finite Element Method in \cite{MvPS04}, an explicit implicit PIDE-FFT method for general L\' evy processes in \cite{CV05} or Predictor Corrector methods to improve the accuracy of the PIDE in \cite{BM06}.
In the best case, all of these methods lead to a time of calibration of roughly one minute (see \cite{AA00}). Can we reduce this computational time? Is it possible to reach a time of calibration as short as the computational time of a closed formula such as Merton's \cite{Mer76}? Our present research responds positively to the above questions.

In order to handle even more general situations we consider, for the one dimensional underlying state process, the solution of the stochastic differential equation (SDE):
\begin{equation}\label{withoutperturbationsde}
dX_{t}=\sigma(t,X_{t-}) d{W}_{t}+\mu(t,X_{t-}) dt+dJ_{t}, \quad X_{0}=x_{0}.
\end{equation} 
For instance one may think of $(X_{t})_{t}$ as the log asset price. Here $(W_{t})_{0\leq t\leq T}$ is a standard real Brownian motion on a filtered probability space $(\Omega,\mathcal{F},(\mathcal{F}_{t})_{0\leq t\leq T},\mathbb{P})$ with the usual assumption on the filtration $(\mathcal{F}_{t})_{0\leq t\leq T}$ and $(J_{t})_{0\leq t\leq T}$ is a compound Poisson process independent of $(W_{t})_{t}$ defined by: 
$J_{t}=\sum_{i=1}^{N_{t}}Y_{i}$
where $N_{t}$ is a counting Poisson process with constant jump intensity $\lambda $  and $(Y_{i})_{i\in \N^{*}}$ are i.i.d. normal variables with mean $\eta_{J}$ and volatility $\gamma_{J}$. Our main objective is to give an accurate analytic approximation of the expected payoff (or fair price of this option) 
\begin{equation*}
\E(h(X_{T}))
\end{equation*} 
for a given terminal function $h$ and for a fixed maturity $T$.

The approximation can be applied to the following models:
\begin{example}\label{LWAA}\textit{AA model on the log-asset.}\\
In this case, $(X_{t})_{t}$ is the logarithm of the underlying asset, $\sigma(t,x)$ is its local volatility and $\mu(t,x)=\lambda(1-e^{\eta_{J}+\frac{\gamma_{J}^{2}}{2}})-\frac{\sigma^{2}(t,x)}{2}$ in order to guarantee the martingale property for $(e^{X_{t}})_{t}$. For a call exercised at maturity $T$, with strike $K $, $h(x)=e^{-\int_{0}^{T}r(u)du}(e^{\int_{0}^{T}(r(u)-q(u))du} e^{x}-K)_{+}$ where $r$ is the deterministic risk-free rate term and $q$ is the deterministic dividend term. This model was derived in \cite{AA00}. In this work we mainly focus our discussion on this model.
\end{example} 
\begin{example}\label{NWAA}\textit{Jump diffusion model on the asset.}\\
 $(X_{t})_{t}$ is the forward contract with maturity $T$, $\sigma(t,x)$ is its volatility and $\mu(t,x)=-\lambda \eta_{J} $.  For a call exercised at maturity T, $h(x)=e^{-\int_{0}^{T}r(u)du}(x-K)_{+}$. The primary focus of this model is the implied normal volatility instead of standard implied Black-Scholes volatility (Japanese markets in \cite{HKLW02}) and it includes the presence of price jumps.
\end{example}

\paragraph{\bf Heuristics of our approximation and model proxy.} In practice, at first glance, it is reasonable to think that $(X_t)_t$ (in the AA model) is approximated by a Merton model, where the coefficients $\mu$ and $\sigma$ only depend on time. We denote this proxy by $(X^M_t)_t$ and it is defined by 
\begin{equation}\label{eq:xm}
dX^M_{t}=\sigma(t,x_0) d{W}_{t}+\mu(t,x_0)dt+dJ_{t}, \quad X^M_{0}=x_{0}.
\end{equation} 
This approximation can be justified by one of the following situations.
\begin{itemize}
\item [\em i)] The functions $\mu(\cdot)$ and $\sigma(\cdot)$ have small variations, which means that $\sigma(t,X_{t-})\approx\sigma(t,x_0)$ and analogously for $\mu$.
\item [\em ii)] The diffusion component is small (i.e. $|\mu|_\infty+|\sigma|_\infty$ is small) and the jump component as well (i.e. $\lambda(|\eta_J|+\gamma_J)$ is small, meaning that the jump frequency or the jump size is small), which results in $X_{t}\approx x_0$. This case is not equivalent to situation {\em i)} because the functions may be small and yet have large variations.
\item [\em iii)] Another obvious reason may be that the maturity $T$ is small, leading to  $X_{T}\approx x_0$.
\end{itemize}
The heuristics {\em i)} and {\em ii)} are coherent with the parameter values taken in \cite{AA00}. When the three conditions are carried out at the same time, we expect our approximation to become even more accurate. Note also that no jump cases ($\lambda=0$) are allowed. The above qualitative features {\em i)} and {\em ii)} are encoded into quantitative constants $M_0$, $M_1$ and $M_J$ defined in (\ref{eq:M}) and will be discussed later in this work. The above heuristic rule implies that
\begin{equation*}
\E(h(X_{T}))=\E(h(X^M_{T}))+{\rm error}.
\end{equation*} 
The term $\E(h(X^M_{T}))$ is the price in the Merton proxy which is explicit (see Remark \ref{remark:merton}). But this sole approximation is too rough to be sufficiently accurate. Our work consists of deriving correction terms for the above equality to attain a remarkably good approximation. 

\paragraph{\bf Smart expansion.} To perform a rigorous analysis, we use a suitable parameterization w.r.t. $\epsilon \in [0,1]$:
\begin{equation}\label{perturbationsde}
dX^{\epsilon}_{t}=\epsilon(\sigma(t,X^{\epsilon}_{t-}) d{W}_{t}+\mu(t,X^{\epsilon}_{t-}) dt+dJ_{t}), X^{\epsilon}_{0}=x_{0},
\end{equation} 
so that $X^{1}_{t}=X_{t}$. We write
\begin{equation}\label{cost}
g(\epsilon)=\E(h(X^{\epsilon}_{T}))
\end{equation} 
and our approach consists of expanding the price (\ref{cost}) with respect to $\epsilon$. But the accuracy of the expansion is not related to $\epsilon$ because the value of interest $\epsilon=1$ is not small. This is a significant difference as compared with singular perturbation techniques. Parameterization is just a tool to derive convenient representations. By using an asymptotic expansion in the context of small diffusions and small jumps (relative to the frequency or to the size), we can establish estimates of the derivatives. This allows us to make an explicit contribution at given order and to control the error. This is achieved by using the infinite dimensional analysis of Malliavin calculus. Here, we focus our analysis on the first terms\footnote{in the former version of this work, terms at any order have been analyzed.}, for which we provide explicit formulas. We also give explicit upper bounds of the errors for general forms of $\mu(\cdot)$ and $\sigma(\cdot)$. However, the smaller the parameters $\mu(\cdot),\sigma(\cdot)$ and $\lambda(|\eta_{J}|+\gamma_{J})$ are, the smaller the maturity $T$ is, or the smaller the derivatives of the functions $\sigma(\cdot)$ and $\mu(\cdot)$ w.r.t. the second variable are, the more accurate the expansion is. Given realistic parameters, the accuracy is indeed very good (less than \textbf{2bp} in implied volatilities for various strikes and maturities). As a result of these expansions, we prove that the price (\ref{cost}) in our general model (\ref{withoutperturbationsde}) equals the price in the Merton model plus a combination of Greeks (still in the Merton model). Hence, numerical evaluation of all these terms is straightforward, with a computational cost equivalent to the closed Merton formula. The residual terms (otherwise stated as error) are also estimated and their amplitudes depend on the smoothness of the payoff. We distinguish three cases: smooth, vanilla (call, put) and binary payoffs. In practice, the vanilla case is likely to be the most useful. 

This is our main contribution. Furthermore, from
 the approximation price we observe that one may obtain a volatility smile for short maturities (since we use the Merton model as a proxy) and a volatility skew for long maturities (due to local volatility function).

\paragraph{\bf Comparison with the literature.} We refer in particular to Hagan et al. in \cite{HKLW02} for the SABR model, to Fouque et al. in \cite{FPS00} for stochastic volatility models, or to Antonelli-Scarletti in \cite{AS07}. In all these works, as opposed to our approach, a perturbation analysis w.r.t. the volatility, the mean reversion parameters, or the correlation, is performed and this leads to writing the price as a main term (essentially a Black-Scholes price) plus an integral of Greeks over maturities. In the time homogeneous case, the authors successfully compute or approximate this integral, which strongly relies on PDE arguments. In our case, we do not approximate the underlying PDE (or the related operator) but owing to Malliavin calculus, we directly focus on the law of the random variable $X_{T}$ given $X_{0}=x_{0}$ and not necessarily on the process for any initial condition. Thus, we are able to handle time inhomogeneous coefficients and jumps as well, without extra effort. This is a very significant difference from previous research.\\   

\paragraph{\bf Outline of the paper.} In the following, we present some notations and assumptions that will be used throughout the paper. Section \ref{sectSTD} is aimed at presenting our methodology in an heuristic way to approximate the expected cost. Rigorous results are proved in Section \ref{sectIP}. In Section \ref{sectFC}, we derive financial modeling consequences from these formulas. These observations lead to justifying simplified choices of the local volatility (of the CEV type), to predict the form of all attainable smiles with their dynamics. In Section \ref{sectNE}, we first give a methodology for implementing the approximation formula. Secondly, we show how to efficiently use our formula for calibrating the model using a relevant algorithm. Finally, we detail numerical applications in calibration for real market data using our simplified form of local volatility. In Section \ref{sectIP}, we analyze the amplitude of the correction and error terms of the approximation formula; the analysis depends on the kinds of payoff (smooth payoff in Theorem \ref{OrderTheoremReg}, vanilla options in Theorem \ref{OrderTheoremReg1}, binary options in Theorem \ref{OrderTheoremReg0}). % and we discuss the impact of the payoff on the accuracy. The proofs of Theorems \ref{OrderTheoremReg}-\ref{OrderTheoremReg1}-\ref{OrderTheoremReg0} are postponed to Section \ref{sectPro}. 
In Appendix \ref{sectApp}, we bring together useful results to make our ``smart expansion'' explicit.

\paragraph{\bf Notations used throughout the paper.}~\\
\textbf{Differentiation.} If these derivatives have a meaning, we write:
\begin{itemize}
\item $\psi^{(i)}_{t}(x)={\frac{\partial^{i} \psi}{\partial x^{i}}}(t,x)$ for every function $\psi$ of two variables.
\item $  X_{i,t}={\frac{\partial^{i} X^{\epsilon}_{t}}{\partial \epsilon^{i}}|_{\epsilon=0}}$ . These processes play a crucial role in the work that follows.
\item When there is no ambiguity, we simply write $\sigma_{t}=\sigma(t,x_{0}),\mu_{t}=\mu(t,x_{0}),\sigma^{(i)}_{t}={\frac{\partial^{i} \sigma}{\partial x^{i}}}(t,x_{0}),\mu^{(i)}_{t}={\frac{\partial^{i} \mu}{\partial x^{i}}}(t,x_{0})$.
\end{itemize}

The following definition is used to distinguish the payoff functions $h$.
\begin{definition}
As per usual, we define $\C_{0}^{\infty}(\R)$ as the space of real infinitely differentiable functions $h$ with compact support (smooth payoffs). The sup-norm of the function $h$ is denoted by $|h|_\infty$. We define $\Hh$ as the space of functions with growth being at most exponential. In other words, a function $h$ belongs to $\Hh$ if $|h(x)|\leq c_{1}e^{c_{2}|x|}$ for any $x$, for two constants $c_{1}$ and $c_{2}$.
\end{definition}
The following notation provides a convenient representation of the correction terms.
\begin{definition}\textbf{Greeks.}\label{Grkdef} Let $Z$ be a random variable. Given a payoff function $h$, we define the $i^{th}$ Greek for the variable $Z$ by the quantity (when it has a meaning) :
\begin{equation*}
\G^{h}_{i}(Z)= \frac{\partial^{i}\E[h(Z+x)]}{\partial x^{i}}\big|_{x=0}.
\end{equation*}
Given appropriate smoothness assumptions concerning $h$, one also has 
\begin{equation*}
\G^{h}_{i}(Z)= \E[h^{(i)}(Z)].
\end{equation*}
\end{definition}

\paragraph{\bf Assumptions.} In order to get accurate approximations, we may assume that coefficients $\sigma$ and $\mu$ are smooth enough.
\begin{itemize}
\item \textit{ \textbf{Assumption} ($R_{4}$). The functions $\sigma(\cdot)$ and $\mu(\cdot)$ are continuously differentiable w.r.t. $x$ up to order $4$. In addition, these functions and their derivatives are uniformly bounded.}
\end{itemize}
The functions and their derivatives could be piecewise continuous w.r.t. the time variable, without changing the following approximation formulas and the following error bounds.

The assumption $(R_{4})$ seems to be restrictive because one requires $\sigma(\cdot),\mu(\cdot)$ and their derivatives w.r.t. $x$ to be bounded. On the one hand, this hypothesis is clearly too strong for us to use in the derivation of our smart expansion: indeed, the reader may check that polynomial growth conditions are sufficient for this purpose. On the other hand, assuming that the derivatives are bounded is much more convenient for explanation purposes. It enables us to state all our error estimates purely in terms of the following constants:
\begin{equation}
\label{eq:M}
\left\{\begin{array}{cl}
M_{1}=&\max_{1\le i\le 4}(|\sigma^{(i)}|_{\infty}+|\mu^{(i)}|_{\infty}),\\
M_{0}=&\max_{0\le i\le 4}(|\sigma^{(i)}|_{\infty}+|\mu^{(i)}|_{\infty}),\\
M_J=&|\eta_{J}|+\gamma_{J}.
\end{array}\right.
\end{equation}
$M_1$, $M_0$ and $M_J$ play complementary roles. 
\begin{itemize}
\item [a)] The constant $M_1$ is a measure of the norm of the derivatives (w.r.t. $x$) of the objective functions $\sigma(\cdot)$ and $\mu(\cdot)$. All our error estimates (see Theorems \ref{OrderTheoremReg}-\ref{OrderTheoremReg1}-\ref{OrderTheoremReg0}) are linear w.r.t. $M_1$, which corroborates the proxy intuition explained in item {\em i)}. The smaller the value of $M_1$ is, the closer $X$ and $X^M$ are, and as a result, approximation is increasingly accurate. At the limit $M_1=0$, the initial model and the proxy coincide ($X_t=X^M_t$) and our approximation formula becomes exact.
\item [b)] The constants $M_0$ and $M_J$ also include estimates of the amplitudes of $\sigma(\cdot), \mu(\cdot)$ and of the jump components. All our error estimates also depend on powers of $M_0$ and $M_J$. This mathematically justifies proxy intuition {\em ii)}. The smaller $M_0$ and $M_J$ are, the better the resulting accuracy.
\end{itemize}
In our next theorems, we also clarify the dependence of our estimates regarding jump frequency $\lambda$ and maturity $T$, because as these parameters decrease, the approximation becomes increasingly accurate.

To perform the infinitesimal analysis, we rely on smoothness properties which are not provided by the payoff functions, but rather by the law of the underlying stochastic models (this is related to Malliavin calculus). The following ellipticity assumption on volatility combined with $(R_{4})$ guarantees these smoothness properties.
\begin{itemize}
\item \textit{ \textbf{Assumption} (E). $\sigma$ does not vanish and for a positive constant $C_E$, one has
\begin{equation*}
 1\leq \frac{|\sigma|_{\infty}}{\sigma_{inf}}\leq C_E
\end{equation*}
where $\sigma_{inf}=\mathop{\inf }_{(t,x)\in \R^{+}\times\R} \sigma(t,x)$.
}
\end{itemize}
We also need to separate our analysis according to payoff smoothness. We thus divide our analysis into three cases.
\begin{itemize}
\item \textit{\textbf{Assumption} ($H_{1}$). $h$ belongs to $\C_{0}^{\infty}(\R)$. This case corresponds to smooth payoffs.}
\item \textit{\textbf{Assumption} ($H_{2}$). $h$ is almost everywhere differentiable. In addition, $h$ and $h^{(1)}$ belong to $\Hh$. This case corresponds to vanilla options (call, put).}
\item \textit{ \textbf{Assumption} ($H_{3}$). $h$ belongs to $\Hh$. This case includes binary options (digital).}
\end{itemize}
\section{Smart Taylor Development}\label{sectSTD}
In this section, we formally show how to replace the price $\E(h(X_T))$ by using that found in the Merton model $\E(h(X^M_T))$ with appropriate correction terms. Rigorous justification of the following expansions is postponed to Section \ref{sectIP}.

The initial trick of our smart expansion lies in the use of the parameterized process $(X^\epsilon_t)_t$ for $\epsilon\in[0,1]$, defined in (\ref{perturbationsde}). Under assumption $(R_{4})$, almost surely for any $t$, $X^{\epsilon}_{t}$ is $C^{3}$ w.r.t $\epsilon$ (see Theorem 2.3 in \cite{FujKun85}). If we put $  X^{\epsilon}_{i,t}={\frac{\partial^{i} X^{\epsilon}_{t}}{\partial \epsilon^{i}}}$, we get 
\begin{align*}
dX^{\epsilon}_{1,t}=&\sigma_t(X^{\epsilon}_{t-}) d{W}_{t}+\mu_t(X^{\epsilon}_{t-}) dt+dJ_{t} \\&+
\epsilon X^{\epsilon}_{1,t-}(\sigma^{(1)}_{t}(X^{\epsilon}_{t-}) d{W}_{t}+\mu^{(1)}_{t}(X^{\epsilon}_{t-}) dt),\quad X^{\epsilon}_{1,0}=0.
\end{align*}
From the definitions, $  X_{i,t}\equiv{\frac{\partial^{i} X^{\epsilon}_{t}}{\partial \epsilon^{i}}|_{\epsilon=0}}$,  $\sigma^{(i)}_{t}\equiv\sigma^{(i)}(t,x_{0})$ and $\mu^{(i)}_{t}\equiv\mu^{(i)}(t,x_{0})$, we easily get
\begin{align*}
dX_{0,t}=&0,\quad X_{0,0}=x_0,\\
dX_{1,t}=&\sigma_{t}dW_{t}+\mu_{t}dt+dJ_{t},\quad X_{1,0}=0,\\
dX_{2,t}=&2X_{1,t-} (\sigma^{(1)}_{t}dW_{t}+\mu^{(1)}_{t}dt),\quad X_{2,0}=0.
\end{align*}
Thus, the Merton model is obtained by the first order expansion of $X^\epsilon$ at $\epsilon=0$:
$$X_{0,T}+X_{1,T}=x_0+X_{1,T}=X^M_T.$$
We now use the Taylor formula twice: first, for $X^{\epsilon}_{T}$ at the second order w.r.t $\epsilon$ around $\epsilon=0$, second for smooth function $h$ at the first order w.r.t $x$ around $x_{0}+X_{1,T}=X^M_T$. One gets:
\begin{align*}
\E[h(X^{1}_{T})]=&\E[h(x_{0}+X_{1,T}+\frac{X_{2,T}}{2}+\cdots)]= \E[h(X^M_T)]+\E[h^{(1)}(X^M_T)\frac{X_{2,T}}{2}]+\cdots.
\end{align*}
Then, the price $\E[h(X_{T})]$ can be approximated by a summation of two terms :
\begin{itemize}
\item $\E[h(X^M_T)]$: The leading order which corresponds to the Merton price (BS price when $\lambda=0$) for the payoff $h$.
\item $\E[h^{(1)}(X^M_T)\frac{X_{2,T}}{2}]$: The correction term which is made explicit in the next theorem. 
\end{itemize}
\begin{theorem}\label{secondorder}(\textbf{Main approximation price formula}).\\
Suppose that the process data fulfills $(R_{4})$ and $(E)$ and that the payoff function fulfills one of the assumptions $(H_{1})$, $(H_{2})$ or $(H_{3})$. Then 
\begin{align}
\E[h(X_{T})]=&\E[h(X^M_T)]+\sum_{i=1}^{3}\alpha_{i,T}\G^{h}_{i}(X^M_T)+\sum_{i=1}^{3}\beta_{i,T}\G^{h}_{i}(X^M_T+Y')+{\rm Error},\label{approxfor}
\end{align}
where
\begin{align*}
\alpha_{1,T}=&\int_{0}^{T} \mu_{t} (\int_{t}^{T}\mu^{(1)}_{s}ds)dt,\\
\alpha_{2,T}=&\int_{0}^{T} ( \sigma^{2}_{t}(\int_{t}^{T}\mu^{(1)}_{s}ds)+ \mu_{t} (\int_{t}^{T}\sigma_{s}\sigma^{(1)}_{s}ds))dt,\\
\alpha_{3,T}=&\int_{0}^{T} \sigma^{2}_{t}  (\int_{t}^{T}\sigma_{s}\sigma^{(1)}_{s}ds)dt,\\
\beta_{1,T}=&\lambda \eta_{J}\int_{0}^{T} t\mu^{(1)}_{t}dt,\\
\beta_{2,T}=&\lambda \int_{0}^{T}t(\gamma_{J}\mu^{(1)}_{t}+\eta_{J}\sigma_{t}\sigma^{(1)}_{t})dt,\\
\beta_{3,T}=&\lambda\gamma_{J}\int_{0}^{T}t\sigma_{t}\sigma^{(1)}_{t}dt,
\end{align*}
 $Y'$ is an independent copy of the variables $(Y_{i})_{i\in \N^{*}}$.\\ In addition, estimates for the error term ${\rm Error}$ in the cases $(H_{1})$, $(H_{2})$ and $(H_{3})$ are respectively given in Theorems \ref{OrderTheoremReg}, \ref{OrderTheoremReg1} and \ref{OrderTheoremReg0}.
\end{theorem}
To prove Theorem \ref{secondorder}, it remains to show that $\E[h^{(1)}(X^M_T)\frac{X_{2,T}}{2}]$ is equal to the two summations of (\ref{approxfor}). The reader familiar with Malliavin calculus for the computations of Greeks (see \cite{FLLLT99}, \cite{Gob04}, \ldots) may recognize in the expansion of $\E[h^{(1)}(X^M_T)\frac{X_{2,T}}{2}]$ the generic form of some derivatives (or Greeks) of $\E[h^{(1)}(X^M_T)]$, derivatives which are written as the expectation of $h^{(1)}(X^M_T)$ multiplied by random weights. This is indeed our methodology to explicitly compute the correction terms in the formula (\ref{approxfor}). 
\begin{proof}
Define the new function $G$ by
$G(x)=h(x+x_{0}+\int_{0}^{T} \mu_{t} dt)$. One has:
\begin{align*}
\E[\frac{X_{2,T}}{2}h^{(1)}(X^M_T)]&=\E[\frac{X_{2,T}}{2}G^{(1)}(\int_{0}^{T}\sigma_{t}d{W}_{t}+J_{T})]\\
&=\E[(\int_{0}^{T} X_{1,t-} (\sigma^{(1)}_{t} d{W}_{t}+\mu^{(1)}_{t} dt)) G^{(1)}(\int_{0}^{T}\sigma_{t}d{W}_{t}+J_{T})].
\end{align*}
Write $(X^{c}_{1,t})_t$ for the continuous part of $(X_{1,t})_t$.
Using Lemma \ref{lemma2} (since $J_{T}$ is independent of $(W_{t})_{t \in [0,T]}$) and ${\rm Leb}\{t\in [0,T]: X_{1,t}=X_{1,t-}\}=0$ a.s. (see page 6 in \cite{Sat99}), one gets:
\begin{align*}
\E[\frac{X_{2,T}}{2}h^{(1)}(X^M_T)]&=\E[(\int_{0}^{T} \sigma_{t}\sigma^{(1)}_{t}X^{c}_{1,t}  dt) G^{(2)}(\int_{0}^{T}\sigma_{t}d{W}_{t}+J_{T})]\\&+\E[(\int_{0}^{T} \mu^{(1)}_{t} X^{c}_{1,t}  dt) G^{(1)}(\int_{0}^{T}\sigma_{t}d{W}_{t}+J_{T})]\\
&+\E[(\int_{0}^{T}  \sigma_{t}\sigma^{(1)}_{t} J_{t}  dt) G^{(2)}(\int_{0}^{T}\sigma_{t}d{W}_{t}+J_{T})]\\
&+\E[(\int_{0}^{T} \mu^{(1)}_{t} J_{t}  dt) G^{(1)}(\int_{0}^{T}\sigma_{t}d{W}_{t}+J_{T})].
\end{align*}
Apply Lemmas \ref{corollary2} and \ref{LJKcalculus} and use Definition \ref{Grkdef} of Greeks to get the result. \qed \end{proof}
\begin{remark}
\label{remark:merton}
 The above price approximation is a summation of three terms: 
\begin{enumerate}
\item $\E[h(X^M_T)]$: The leading order corresponding to the price when the functions $(\sigma_t)_t$ and $(\mu_t)_t$ are deterministic. We know that in this case, there is a closed formula : either the Merton closed formula for call (put), or FFT tools for any other payoff because the characteristic function of $X^M_{T}$ is explicit. For instance, the formula for a call in the Merton model (see \cite{Mer76}) on the log asset is:
\begin{align*}
\sum_{i=0}^{\infty} \frac{(\lambda T)^{i}}{i!}e^{-\lambda T-\int_{0}^{T}r(u)du}BSCall\bigg( F_{T}e^{i(\eta_{J}+\frac{\gamma_{J}^{2}}{2})},K,T,
\sqrt{\frac{\int_{0}^{T}\sigma^{2}_{t}dt +i\gamma_{J}^{2}}{T}}\bigg),
\end{align*} 
where
\begin{align*}
F_{T}=e^{x_{0}+\int_{0}^{T}(r(u)-q(u))du+ \lambda(1-\exp(\eta_{J}+\gamma_{J}^{2}/2))T},
\end{align*}
and $BSCall(S,K,T,v)$ is the Black-Scholes price for a call on an underlying $S_{t}$ with initial condition $S_{0}=S$, volatility $v$, exercised at maturity $T$ and strike $K$, where the risk-free rate and the dividend yield are set at $0\%$. 
\item $\sum_{i=1}^{3}\alpha_{i,T}\G^{h}_{i}(X^M_T)$: The volatility and drift correction term which depends on the first derivatives of $\mu$ and $\sigma$. This term can be computed as easily as the main term. 
\item $\sum_{i=1}^{3 }\beta_{i,T}\G^{h}_{i}(X^M_T+Y')$: The jump correction term which depends on the first derivatives of $\mu$, $\sigma$ and on the jump parameters. Since $Y'$ is also Gaussian and independent of $X^M_{T}$, the computation of these Greeks are similar to the previous ones, by adding to the mean $\int_{0}^{T}\mu_{t}dt$ and variance $\int_{0}^{T}\sigma_{t}^{2}dt$ the quantities $\eta_{J}$ and $\gamma^{2}_{J}$.
\end{enumerate}
\end{remark}
\begin{remark}
In the AA model on the log-asset, one has:
\begin{align*}
\alpha_{1,T}   =&\frac{1}{2}\int_{0}^{T}  \sigma^{2}_{t} (\int_{t}^{T}\sigma_{s}\sigma^{(1)}_{s}ds)dt+ \lambda ( e^{\eta_{J}+\frac{\gamma^{2}_{J}}{2}}-1) \int_{0}^{T} t\sigma_{t}\sigma^{(1)}_{t}dt,\\
\alpha_{2,T}   =&-\frac{3}{2}\int_{0}^{T}  \sigma^{2}_{t}(\int_{t}^{T}\sigma_{s}\sigma^{(1)}_{s}ds)dt- \lambda ( e^{\eta_{J}+\frac{\gamma^{2}_{J}}{2}}-1) \int_{0}^{T} t\sigma_{t}\sigma^{(1)}_{t}dt,\\
\alpha_{3,T}   =&\int_{0}^{T} \sigma^{2}_{t}  (\int_{t}^{T}\sigma_{s}\sigma^{(1)}_{s}ds)dt,\\
\beta_{1,T}=&-\lambda \eta_{J}\int_{0}^{T} t\sigma\sigma^{(1)}_{t}dt,\\
\beta_{2,T}=&\lambda (\eta_{J}-\gamma_{J})  \int_{0}^{T}t\sigma_{t}\sigma^{(1)}_{t}dt,\\
\beta_{3,T}=&\lambda\gamma_{J}\int_{0}^{T}t\sigma_{t}\sigma^{(1)}_{t}dt.
\end{align*}
Thus, the computation of these constants is simply reduced to that of $\int_{0}^{T}t\sigma_{t}\sigma^{(1)}_{t}dt$ and $\int_{0}^{T} \sigma^{2}_{t} (\int_{t}^{T}\sigma_{s}\sigma^{(1)}_{s}ds)dt$. 
\end{remark}
We note that we can perform higher order approximation formulas that remain explicit. The only difference is that the number of random variables used as arguments for the Greeks will increase with each order, and it is within the set $(X_{1,T}+Y'_{1}+\cdots+Y'_{i})_{i \in  \N}$. We refer to \cite{Miri} for higher order terms.
\section{Financial Modeling Consequences}\label{sectFC}
For simplicity, we consider the AA model on the log-asset (an analogous statement would be available for the jump diffusion model on the asset). 

The standard Gaussian framework as developed by Black-Scholes (1973) and Merton (1976) is realized by choosing a constant volatility function $\sigma(\cdot)$ (the computation is still possible for a function dependent only on time). In order to arrive at a coherent, appropriate analysis and modelling for a fixed income market (without jump)  Andersen and Andreasen \cite{AA02} take  a parametric form for $\sigma$:
\begin{equation}\label{SFVol}
\sigma(t,x)=\nu(t) e^{(\beta(t)-1)x}, 
\end{equation}
where $\nu(t)$ the relative volatility function, $\beta(t)$ is a time-dependent constant elasticity of variance (CEV). Piterbarg\footnote{If $\sigma_{Pit} $ is the local volatility used in \cite{Pit05} and  $L(t)=e^{\int_{0}^{t}(r(u)-q(u))du}$, one has $\sigma(t,x)=\sigma_{Pit}(t,L_{t}e^{x})$.} \cite{Pit05} uses the same form but applies it to Power Reverse Dual Currency swaps in order to handle the skew for the FX. 

Because of $\mu(t,x)=\lambda(1-e^{\eta_{J}+\frac{\gamma_{J}^{2}}{2}})-\frac{\sigma^{2}(t,x)}{2}$, the approximation formula (\ref{approxfor}) depends only on $\sigma(t,x_{0}),\sigma^{(1)}(t,x_{0}),\lambda,\eta_{J}$ and $\gamma_{J}$.
The volatility given in equation (\ref{SFVol}) may generate all possible values of the following time-dependent functions $\sigma(t,x_{0})=\nu(t)e^{(\beta(t)-1)x_{0}}$ and $\sigma^{(1)}(t,x_{0})=(\beta(t)-1)\nu(t) e^{(\beta(t)-1)x_{0}}$, because it has two degrees of freedom $\nu(t)$ and $\beta(t)$. So this kind of volatility potentially creates all attainable prices in this class of models, and thus all attainable Black-Scholes smiles. This justifies interest in CEV-type volatility (\ref{SFVol}).  

\textbf{Attainable Black-Scholes smiles using the model.} Can we predict the general form of the smiles generated by this model?
\begin{itemize}
\item For short maturity: using our approach, the model is close to the Merton model related to $X^M_T$. Therefore, the shape of implied volatilities forms a smile centered on a point close to the money, which is on the left when $\eta_{J}+\frac{\gamma_{J}^{2}}{2}>0$ (on the right when $\eta_{J}+\frac{\gamma_{J}^{2}}{2}<0$ ).\\
\textit{Formal Proof: Using the approximation formula, the correction terms are $O(T)$. So when $T$ decreases to zero, the price converges to the Merton price. The second statement is easy to check. One can follow the approach of \cite{Gat02,Mat00} using characteristic functions, or can prove it directly using some derivations of the Merton formula \cite{Mer76}.}
\item For long maturity: the smile becomes a skew which is due to the local volatility function (because the smile for the Merton model flattens for long maturity).
\end{itemize}

\textbf{Smile Dynamics.} The model has the Merton model as a good proxy. The implied volatilities for the Merton model are increasing and depend only on the ratio between the forward and the strike. Therefore, the smile should move in the same direction as the forward.
\section{Numerical Experiments}\label{sectNE}
In this section, we give details of the implementation for the approximation (\ref{approxfor}) and illustrate the accuracy of our formula. After that, a generic bootstrap algorithm  for calibration purposes is derived. Finally, a numerical application of this algorithm is applied to market data (currency options).
\subsection{Numerical Implementation}\label{subNI}
The case of time homogeneous parameters $\sigma_t,\sigma^{(1)}_{t},\mu_t$ and $\mu^{(1)}_{t}$ gives us the coefficients $\alpha$ and $\beta$ exactly (see their expressions in Theorem \ref{secondorder}).\\% This result is still valid for higher orders.\\
In addition, when these parameters are time-dependent, there are two cases.
\begin{itemize}
\item Either the data are smooth. In this case, we use a Gauss-Legendre quadrature formula (see \cite{PTVF}) for the calculation of the coefficients $\alpha$ and $\beta$.
\item Or the data are piecewise constant. In this case, we can give explicit expressions of $\alpha$ and $\beta$ in terms of the piecewise constant data. Let $T_{0}=0\leq T_{1}\leq \cdots\leq  T_{n}=T$ such that $\sigma_t,\sigma^{(1)}_{t},\mu_t$ and $\mu^{(1)}_{t}$ are constant at each interval $]T_{i},T_{i+1}]$ and are equal respectively to $\sigma_{T_{i+1}},\sigma^{(1)}_{T_{i+1}},\mu_{T_{i+1}}$ and $\mu^{(1)}_{T_{i+1}}$.
Before giving the recursive formula, we need to introduce the following functions: 
${\omega}_{1,t}=\int_{0}^{t} \sigma^{2}_{s}ds ,{\omega}_{2,t}=\int_{0}^{t} \mu_{s}ds $.
\end{itemize}
\begin{proposition}\label{Reccal}\textbf{Recursive formula.}\\
For piecewise constant coefficients, one has:
\begin{align*}
\alpha_{1,T_{i+1}}=&\alpha_{1,T_{i}}+(T_{i+1}-T_{i}) \mu^{(1)}_{T_{i+1}} {\omega}_{2,T_{i}}+\frac{(T_{i+1}-T_{i})^{2}}{2} \mu_{T_{i+1}}\mu^{(1)}_{T_{i+1}}, \\
\alpha_{2,T_{i+1}}=&\alpha_{2,T_{i}}+(T_{i+1}-T_{i}) (\mu^{(1)}_{T_{i+1}} \omega_{1,T_{i}}+\sigma_{T_{i+1}}\sigma^{(1)}_{T_{i+1}} {\omega}_{2,T_{i}})\\+&\frac{(T_{i+1}-T_{i})^{2}}{2} (\sigma^{2}_{T_{i+1}}\mu^{(1)}_{T_{i+1}}+ \mu_{T_{i+1}}\sigma_{T_{i+1}}\sigma^{(1)}_{T_{i+1}}),\\
\alpha_{3,T_{i+1}}=&\alpha_{3,T_{i}}+(T_{i+1}-T_{i}) \sigma_{T_{i+1}}\sigma^{(1)}_{T_{i+1}} {\omega}_{1,T_{i}}+\frac{(T_{i+1}-T_{i})^{2}}{2} \sigma^{3}_{T_{i+1}}\sigma^{(1)}_{T_{i+1}},\\
\beta_{1,T_{i+1}}=&\beta_{1,T_{i}}+\lambda \eta_{J} \frac{(T^{2}_{i+1}-T^{2}_{i})}{2} \mu^{(1)}_{T_{i+1}} ,\\
\beta_{2,T_{i+1}}=&\beta_{2,T_{i}}+\lambda  \frac{(T^{2}_{i+1}-T^{2}_{i})}{2} (\gamma_{J}\mu^{(1)}_{T_{i+1}}+\eta_{J} \sigma_{T_{i+1}}\sigma^{(1)}_{T_{i+1}}),\\
\beta_{3,T_{i+1}}=&\beta_{3,T_{i}}+\lambda \gamma_{J} \frac{(T^{2}_{i+1}-T^{2}_{i})}{2} \sigma_{T_{i+1}}\sigma^{(1)}_{T_{i+1}},\\
\omega_{1,T_{i+1}}=&{\omega}_{1,T_{i}}+ (T_{i+1}-T_{i})\sigma^{2}_{T_{i+1}}, \\
\omega_{2,T_{i+1}}=&{\omega}_{2,T_{i}}+(T_{i+1}-T_{i})\mu_{T_{i+1}}.
\end{align*}
\end{proposition}
\begin{proof}
According to Theorem \ref{secondorder}, one has:
\begin{align*}
\alpha_{1,T_{i+1}}=&\int_{0}^{T_{i}} \mu_{t} (\int_{t}^{T_{i+1}} \mu^{(1)}_{s}ds)dt+ \int_{T_{i}
}^{T_{i+1}} \mu_{t} (\int_{t}^{T_{i+1}} \mu^{(1)}_{s}ds)dt\\
=&\alpha_{1,T_{i}}+\int_{0}^{T_{i}} \mu_{t} (\int_{T_{i}}^{T_{i+1}} \mu^{(1)}_{s}ds)dt+ \int_{T_{i}
}^{T_{i+1}} \mu_{t} (\int_{t}^{T_{i+1}} \mu^{(1)}_{s}ds)dt\\
=&\alpha_{1,T_{i}}+ (\int_{T_{i}}^{T_{i+1}} \mu^{(1)}_{s}ds) \int_{0}^{T_{i}} \mu_{t} dt+ \int_{T_{i}
}^{T_{i+1}} \mu_{t} (\int_{t}^{T_{i+1}} \mu^{(1)}_{s}ds)dt\\
=&\alpha_{1,T_{i}}+(T_{i+1}-T_{i}) \mu^{(1)}_{T_{i+1}} {\omega}_{2,T_{i}}+\frac{(T_{i+1}-T_{i})^{2}}{2} \mu_{T_{i+1}}\mu^{(1)}_{T_{i+1}}.
\end{align*}
The other terms are calculated analogously.\qed \end{proof}

\subsection{Accuracy of the approximation}\label{subsubAA}
Here, we give a short example of the performance of our method. The jump parameters have been set to: $\lambda=30\%,\eta_J=-8\%,\gamma_J=35\% $. 
These parameters are not small, especially for the jump intensity $\lambda$ and the jump volatility $\gamma_J$. The piecewise constant functions $\nu$ and $\beta$ defined in (\ref{SFVol}) are equal respectively at each interval of the form $[\frac{i}{20},\frac{i+1}{20}] $
to $25\%-i \times 0.11\%$ and $100\%-i\times 0.75\% $.
The spot, the risk-free rate and the dividend yield are set respectively to $100,4\%$ and $0\%$.

We observe in the table below that the errors of implied Black-Scholes volatilities between our approximation and the price calculated using a PIDE method do not exceed \textbf{2 bp} for a large range of strikes and maturities. The computational time of our formula is less than four milliseconds on a $2.6$ GHz Pentium PC. The accuracy of our formula turns out to be excellent.
 \begin{table}[ht]
\caption{Error in implied Black-Scholes volatilities (in bp=0.01\%) between the approximation formula and the PIDE method expressed as a function of maturities in fractions of years and relative strikes.}
\label{tab:bp}
\begin{tabular}{p{1.5cm}p{1.5cm}p{1.5cm}p{1.5cm}p{1.5cm}p{1.5cm}}
\hline\noalign{\smallskip}
T/K & 70\% & 85\%&100\% & 120\%& 150\%
\\
\noalign{\smallskip}\hline\noalign{\smallskip}
3M& 0.02 & -0.03 &-0.92 & -0.07 & -0.12
\\
1Y & 0.04 & 0.06 & 0.15 & -0.11 &0.01
\\
3Y & 0.22 & -0.23 &0.11 &0.41&0.31
\\
5Y & 1.39 &1.06 & -0.01 & 1.85 &1.76
\\
\noalign{\smallskip}\hline
\end{tabular}
\end{table} 

\subsection{Calibration issues}\label{subAC}
For this kind of model (AA model on the log asset or on the asset itself), calibration is still challenging as this model has no analytical formula. We can still perform a numerical calibration using the forward PIDE as explained in \cite{AA00}, but the time of calibration remains quite long (about one minute). With our approach, we can shorten the duration of calibration to less than one second, because our computation of the model price takes four milliseconds as previously mentioned. We achieve that by a simple bootstrapping algorithm using the path dependent formula.

\paragraph{Bootstrap algorithm for piecewise data .}
Suppose that we want to fit option prices for $n$ maturities $T_{0}=0\leq T_{1}\leq \cdots\leq T_{n}$ and $m$ strikes $ K_{1},\cdots,K_{m}$. 
First, we search the parameters $\lambda,\eta_{J}$ and $\gamma_{J}$ with best fit. At each interval $]T_{i-1},T_{i}]$, the data $\sigma$, $\sigma^{(1)}$, $\mu$ and $\mu^{(1)}$  are constant, equal respectively to $\sigma_{T_{i}}$, $\sigma^{(1)}_{T_{i}}$, $\mu_{T_{i}}$ and $\mu^{(1)}_{T_{i}}$, and depending on the vector $\chi_{i}=(\nu(T_{i}),\beta(T_{i}))$ (see formula \ref{SFVol}).
Starting at $i=1$, we express the coefficients $\alpha_{j,T_i}$ and $\beta_{j,T_i}$ as a function of $\chi_{i}$, recursively using Proposition (\ref{Reccal}). We apply a local minimization algorithm (for instance, the Levenberg-Marquardt as described in \cite{PTVF}) in order to fit the implied volatilities for all strikes $K_{1},\cdots,K_{m}$ at maturity $T_{i}$ using our approximation (\ref{approxfor}). Once the vector $\chi_{i}$ is found, we go to the next step $i+1$, update $\alpha$ and $\beta$ and compute $\chi_{i+1}$.\\

This calibration procedure is not completely safe. Sometimes we encounter instability problems. The final parameters depend on the initial guess. Moreover, there are many local minima. To avoid these problems, we could use a regularization method based on relative entropy (see \cite{cont:tank:03}), but these issues are not in direct relation with the accuracy of our formula. We think that the set of calibrated options (call/put) does not contain enough information on the future volatility to ensure a good calibration. Therefore, it is presumably worth including volatility options in the set of calibrated instruments. This is a topic for further research.

\paragraph{Calibration results.} Here, we calibrate the EUR/USD exchange rate. The surface of implied Black-Scholes volatility is given in table \ref{tab:implvol}.

\begin{table}[ht]
\caption{Implied Black-Scholes volatilities for the EUR/USD rate expressed as a function of maturities in fractions of years and relative strikes. The spot is equal to $1.54$. }
\label{tab:implvol}
\begin{center}\begin{tabular}{c c c c c c}%p{1cm} p{1.5cm} p{1.5cm} p{1.5cm} p{1.5cm} p{1cm}}
\hline\noalign{\smallskip}
   T/K &       92\% &       96\% &      100\% &      108\% \\
\noalign{\smallskip}\hline\noalign{\smallskip}
        6M &    10.82\% &    10.65\% &    10.53\% &    10.56\% \\

        1Y &    10.84\% &    10.70\% &    10.63\% &    10.66\% \\

      1.5Y &    10.71\% &    10.60\% &    10.56\% &    10.58\% \\

        2Y &    10.60\% &    10.48\% &    10.46\% &    10.47\% \\
\noalign{\smallskip}\hline
\end{tabular}\end{center}
\end{table}

The jump parameters for the calibrated model are $\lambda=1.21\%$, $\eta_J=-19.07\%$ and $\gamma_J=40.30\% $. The diffusion parameters $\nu$ and $\beta$ for the calibrated model are given in table \ref{tab:nu,beta}. These values are realistic. The errors between the implied volatilities generated by the calibrated model and the market data are given in table \ref{tab:errorimplvol}.

\begin{table}[ht]
\caption{Calibrated values of the piecewise constant functions $\nu$ and $\beta$.}
\label{tab:nu,beta}
\begin{center}
\begin{tabular}{c c c}
\hline\noalign{\smallskip}
$T$ & $\nu$   & $\beta$ \\
\noalign{\smallskip}\hline\noalign{\smallskip}
6M & 10.31\%& 98.81\%    \\ 
1Y & 10.27\%& 100\%    \\
1.5Y & 9.90\%& 100\%    \\
2Y  & 9.43\% & 100\%    \\
\noalign{\smallskip}\hline
\end{tabular}
\end{center}
\end{table}
\begin{table}[ht]
\caption{Errors between implied Black-Scholes volatilities for the EUR/USD rate and those calculated within the calibrated model (in bp) expressed as a function of maturities in fractions of years and relative strikes. The spot is equal to $1.54$. }
\label{tab:errorimplvol}
\begin{center}\begin{tabular}{c c c c c c}%l p{1.5cm} p{1.5cm}p{1.5cm} p{1.5cm} p{1.5cm}}
\hline\noalign{\smallskip}
  T/K &       92\% &       96\% &      100\% &      108\% \\
\noalign{\smallskip}\hline\noalign{\smallskip}
        6M &         -4 &          3 &         -1 &         -3 \\

        1Y &          2 &          1 &          0 &          2 \\

      1.5Y &         -1 &         -3 &         -2 &          1 \\

        2Y &          2 &         -1 &          1 &          4 \\
\noalign{\smallskip}\hline
\end{tabular}
\end{center}
\end{table}

The errors show that our model is a good model for the FX rate EUR/USD. Within our relevant algorithm, we are able to fit a $4\times 4$ grid of quoted prices in less than \textbf{1 s}.
\section{Error Analysis}\label{sectIP}
This section is devoted to the mathematical justification of Theorem \ref{secondorder} and to the statement and proofs of upper bounds for the error term in (\ref{approxfor}). For this, the analysis differs according to the payoff smoothness (smooth, vanilla or binary). We start with the smooth case (subsection \ref{subSP}), which is less technical. Then, we handle the two other cases (call/put and binary options), which requires the use of Malliavin calculus.

Throughout these computations, we aim at emphasizing the dependence of error upper bounds in terms of: the constants $M_0, M_1$ and $M_J$ defined in (\ref{eq:M}), the jump frequency $\lambda$ and the maturity $T$, in order to support the heuristic choice of the model proxy (see the discussion in the introduction).

\paragraph{\bf Additional notation.} 
\begin{itemize}
\item {\em About floating constants and upper bounds.} 
In the following statements and proofs, for the upper bounds we use numerous constants, that are not relabelled during the computations. We simply use the unique notation
$$A\leq_c B$$
to assert that $A\le c B$, where $c$ is a positive constant depending on the model parameters $M_0$, $M_1$, $M_J$, $\lambda$, $T$, $C_E$ (defined in assumption $(E)$) and on other universal constants. The constant $c$ remains bounded when the model parameters go to 0, and it is uniform w.r.t. the parameter $\epsilon\in[0,1]$. When informative, we make clear the dependence of upper bounds w.r.t. $M_0$, $M_1$, $M_J$, $\lambda$ and $T$.
\item {\em Miscellaneous.} As usual, the $\L_p$-norm of a real random variable $Z$ is denoted by $\|Z\|_p=[\E|Z|^p]^{1/p}$. In the proofs, the derivatives of the parameterized process $X^\epsilon$ are useful: they are defined by
 $  X^{\epsilon}_{i,t}={\frac{\partial^{i} X^{\epsilon}_{t}}{\partial \epsilon^{i}}}$.
\end{itemize}

\subsection{Error analysis for smooth payoff (under ($H_1$))}\label{subSP} 
We begin our error analysis with the case of smooth payoff ($h \in \C^{\infty}_{0} (\R)$).
\begin{theorem}\label{OrderTheoremReg} \textbf{Error for smooth payoff.} Assume that $(R_{4})$ holds and that the payoff function $h$ fulfills Assumption $(H_{1})$. Then the error term in Theorem \ref{secondorder} satisfies the following estimate:
\begin{equation}\label{eq:OrderTheoremReg}
|{\rm Error}|\lc \sup_{j=1,2}|h^{(j)}|_{\infty} (M_1\sqrt T)\big((M_0\sqrt T)^2 +M_J^2\sqrt{\lambda T}\big).
\end{equation}
\end{theorem}
Let us briefly comment on the upper bound, making reference to the introduction. If the functions $\sigma(\cdot)$ and $\mu(\cdot)$ are only time dependent ($M_1=0$), the approximation formula (\ref{approxfor}) is exact (the model and the proxy coincide). If they do not vary much w.r.t. $x$ ($M_1$ is small), the accuracy is still good in view of (\ref{eq:OrderTheoremReg}). If the coefficients $\sigma(\cdot),\mu(\cdot)$ and their derivatives and the jump size parameters are all small, the formula becomes very accurate. For instance, in a multiplicative case where $\sigma(t,x) =\Delta s(t,x)$, $\mu(t,x) =\Delta m(t,x)$ and $|\eta_{J}|+\gamma_{J} \leq \Delta$ for a small parameter $\Delta$, it readily follows that $M_{1},M_{0},M_J=O(\Delta)$. Thus 
$$|{\rm Error}|=O\big(\Delta^3 T [\sqrt T+\sqrt \lambda]\big).$$
Consequently, we may refer to the formula (\ref{approxfor}) in Theorem \ref{secondorder} as an approximation of order 2 w.r.t. the amplitudes of the data (with error terms of order 3).

These features arise similarly for the other examples of payoff smoothness.
\begin{proof} It is divided into several steps. First, we write the SDEs satisfied by the three first derivatives of $X^\epsilon_t$ w.r.t. $\epsilon$. Second, we give tight $\L_p$ upper bounds on these derivatives. Finally, we combine these estimates with our smart expansion to complete the proof of Theorem \ref{OrderTheoremReg}.
\paragraph{Step 1. Differentiation of $X^\epsilon$.} Under $(R_{4})$, almost surely $X^{\epsilon}_{t}$ is $C^{3}$ w.r.t $\epsilon$ for any $t$ (see Theorem 2.3 in \cite{FujKun85}) and the derivatives are obtained by successive differentiations of the initial SDE (\ref{perturbationsde}). Thus, direct computations lead to
\begin{align}\label{eq:EDS1}
dX^{\epsilon}_{1,t}=&\sigma_t(X^{\epsilon}_{t-}) d{W}_{t}+\mu_t(X^{\epsilon}_{t-}) dt+dJ_{t} +
\epsilon X^{\epsilon}_{1,t-}(\sigma^{(1)}_{t}(X^{\epsilon}_{t-}) d{W}_{t}+\mu^{(1)}_{t}(X^{\epsilon}_{t-}) dt),\\
\nonumber dX^{\epsilon}_{2,t}=&[2 X^{\epsilon}_{1,t-}\sigma^{(1)}_t(X^{\epsilon}_{t-})+
\epsilon (X^{\epsilon}_{1,t-})^2 \sigma^{(2)}_{t}(X^{\epsilon}_{t-})]dW_t\\
\nonumber&+[2 X^{\epsilon}_{1,t-}\mu^{(1)}_t(X^{\epsilon}_{t-})+
\epsilon (X^{\epsilon}_{1,t-})^2 \mu^{(2)}_{t}(X^{\epsilon}_{t-})]dt\\
&+\epsilon X^{\epsilon}_{2,t-} (\sigma^{(1)}_{t}(X^{\epsilon}_{t-}) dW_t+
\mu^{(1)}_{t}(X^{\epsilon}_{t-})dt),\label{eq:EDS2}\\
\nonumber dX^{\epsilon}_{3,t}=&[3 X^{\epsilon}_{2,t-}\sigma^{(1)}_t(X^{\epsilon}_{t-})+3(X^{\epsilon}_{1,t-})^2\sigma^{(2)}_t(X^{\epsilon}_{t-})+
3 \epsilon X^{\epsilon}_{1,t-}X^{\epsilon}_{2,t-}\sigma^{(2)}_t(X^{\epsilon}_{t-})\\
\nonumber &+\epsilon (X^{\epsilon}_{1,t-})^3 \sigma^{(3)}_{t}(X^{\epsilon}_{t-})]dW_t+[3 X^{\epsilon}_{2,t-}\mu^{(1)}_t(X^{\epsilon}_{t-})+3 (X^{\epsilon}_{1,t-})^2\mu^{(2)}_t(X^{\epsilon}_{t-})\\
&\nonumber+3\epsilon X^{\epsilon}_{1,t-}X^{\epsilon}_{2,t-}\mu^{(2)}_t(X^{\epsilon}_{t-})+\epsilon (X^{\epsilon}_{1,t-})^3 \mu^{(3)}_{t}(X^{\epsilon}_{t-})]dt\\
&+\epsilon X^{\epsilon}_{3,t-} (\sigma^{(1)}_{t}(X^{\epsilon}_{t-}) dW_t+
\mu^{(1)}_{t}(X^{\epsilon}_{t-})dt)\label{eq:EDS3}.
\end{align}
Their initial conditions are all equal to 0. Notice that unlike $X^\epsilon$ and $X^\epsilon_1$, the processes $X^\epsilon_{2}$ and $ X^\epsilon_{3}$ are continuous.
\paragraph{Step 2. Tight upper bounds.} We aim at proving the following estimates for any $p\ge 2$:
\begin{align}
  \label{eq:x1:Lp}
\E|X^{\epsilon}_{1,t}|^p&\lc (M_0\sqrt T)^p + M_J^p \lambda T,\\
\label{eq:x2:Lp}
\E|X^{\epsilon}_{2,t}|^p&\lc  (M_1\sqrt T)^p\big((M_0\sqrt T)^p + M_J^p \lambda T\big),\\
\label{eq:x3:Lp}
\E|X^{\epsilon}_{3,t}|^p&\lc  (M_1\sqrt T)^p\big((M_0\sqrt T)^{2p} + M_J^{2p} \lambda T\big),
\end{align}
uniformly for $t\le T.$\\ The existence of any moment is easy to establish, but here, we emphasize the dependence of the upper bounds w.r.t. the constants $M_0, M_1, M_J, \lambda$ and $T$. Let us first prove the inequality (\ref{eq:x1:Lp}). From (\ref{eq:EDS1}), apply Lemma \ref{Jlem} to the jump component and Burkholder-Davis-Gundy inequalities to the Brownian part, to deduce
\begin{align*}
\E|X^{\epsilon}_{1,t}|^p\lc & t^{p/2-1}\int_0^t \E|\sigma_s(X^{\epsilon}_{s})|^p ds + t^{p-1}\int_0^t \E|\mu_s(X^{\epsilon}_{s})|^p ds+ M_J^p\lambda t\\ &+ t^{p/2-1}\int_0^t \E|X^{\epsilon}_{1,s}\sigma^{(1)}_{s}(X^{\epsilon}_{s})|^p ds+t^{p-1}\int_0^t \E|X^{\epsilon}_{1,s}\mu^{(1)}_{s}(X^{\epsilon}_{s})|^p ds\\
\lc & T^{p/2} M^p_0+ M_J^p\lambda T+ T^{p/2-1}M_1^p\int_0^t \E|X^{\epsilon}_{1,s}|^p ds.
\end{align*}
Using Gronwall's lemma, we easily complete the proof of (\ref{eq:x1:Lp}). For the second inequality (\ref{eq:x2:Lp}), we proceed analogously and we obtain:
\begin{align*}
\E|X^{\epsilon}_{2,t}|^p\lc &  T^{p/2} M_1^p(\sup_{s\le t}\E|X^{\epsilon}_{1,s}|^p +\sup_{s\le t}\E|X^{\epsilon}_{1,s}|^{2p}).
\end{align*}
Thus, plugging the estimate (\ref{eq:x1:Lp}) into the previous inequality directly leads to (\ref{eq:x2:Lp}). Now let us prove the inequality (\ref{eq:x3:Lp}). As before, apply BDG inequalities combined with Gronwall's lemma to obtain that
\begin{align*}
\E|X^{\epsilon}_{3,t}|^p\lc &  T^{p/2} M_1^p(\sup_{s\le t}\E|X^{\epsilon}_{2,s}|^p +\sup_{s\le t}\E|X^{\epsilon}_{1,s}|^{2p}+\sup_{s\le t}\E|X^{\epsilon}_{1,s}X^{\epsilon}_{2,s}|^{p}+\sup_{s\le t}\E|X^{\epsilon}_{1,s}|^{3p}).
\end{align*}
Use $\E|X^{\epsilon}_{1,s}X^{\epsilon}_{2,s}|^{p}\leq \frac 12 (\E|X^{\epsilon}_{1,s}|^{2p}+\E|X^{\epsilon}_{2,s}|^{2p})$ and the previous inequalities 
(\ref{eq:x1:Lp}-\ref{eq:x2:Lp}). Then bringing together different contributions easily leads to the required estimate (\ref{eq:x3:Lp}).
\paragraph{Step 3. Completion of the proof.} We follow the formal computations done at the beginning of Section \ref{sectSTD}, but more carefully. Let us introduce
\begin{equation}
  \label{eq:bar:x}
  \overline X_{2,T}=\int_0^1 X_{2,T}^\epsilon (1-\epsilon) d\epsilon,\quad \overline X_{3,T}=\int_0^1 X_{3,T}^\epsilon \frac{(1-\epsilon)^2}2 d\epsilon.
\end{equation}
Then applications of Taylor expansions of $X^\epsilon_T$ at $\epsilon=0$ readily give these equalities:
\begin{align*}
  %\label{eq:taylor1}
  X_T=X^M_T+\overline X_{2,T},\quad X_T=X^M_T+\frac 12 X_{2,T}+\overline X_{3,T}
\end{align*}
where we have used $X^M_T=x_0+X_{1,T}$. Thus a second order Taylor expansion of $h$ at point $X^M_T$ writes
\begin{align*}
\E[h(X^{1}_{T})]=&\E[h(X^M_T+\frac{X_{2,T}}{2}+\overline X_{3,T})]\nonumber\\
=&\E[h(X^M_T)]+\E[h^{(1)}(X^M_T)\frac{X_{2,T}}{2}]+\E[h^{(1)}(X^M_T)\overline X_{3,T}]\nonumber\\
&+\int_{0}^{1}\E[h^{(2)}((1-v)X^M_T+v X_T)(\overline X_{2,T})^2](1-v)dv. %\label{TaylorReg}.
\end{align*}
This proves that the Error term in (\ref{approxfor}) for smooth payoff equals
\begin{equation}
  \label{eq:error:smooth}
{\rm Error}=
  \E[h^{(1)}(X^M_T)\overline X_{3,T}]+\int_{0}^{1}\E[h^{(2)}((1-v)X^M_T+v X_T)(\overline X_{2,T})^2](1-v)dv.
\end{equation}
Then it readily follows that
\begin{align*}
|  {\rm Error}|\lc |h^{(1)}|_\infty \sup_{\epsilon\in[0,1]} (\E|X_{3,T}^\epsilon|^{2})^{\frac{1}{2}} +|h^{(2)}|_\infty \sup_{\epsilon\in[0,1]}\E|X_{2,T}^\epsilon|^2.
\end{align*}
%check passage form 1 to 2 in L_{p} norm for $X^{\epsilon}_{3,t}$.
It is now straightforward to obtain Theorem \ref{OrderTheoremReg}, by using estimates (\ref{eq:x2:Lp}-\ref{eq:x3:Lp}) with $p=2$.
\qed \end{proof}
A careful inspection of the previous proof shows that assumption $(R_3)$ is sufficient to derive the error estimate (\ref{eq:OrderTheoremReg}).

\subsection{Error analysis for vanilla payoff  (under ($H_2$))}\label{subVA} 
This case has practical importance, because it includes call/put options. Regarding the error estimates related to Theorem \ref{secondorder}, we have paved the way with the case of smooth payoff. Nevertheless, there are some technical differences. The main one is that our previous proof represents the error in terms of the second derivative of the payoff, which is meaningless here. The additional ingredient is the Malliavin calculus integration by parts formula to avoid this second derivative appearing. We now state our main result when the payoff $h$ is almost everywhere differentiable (with sub-exponential growth conditions).
\begin{theorem}\label{OrderTheoremReg1} \textbf{Error for vanilla payoff.} Assume that $(R_{4})$ and $(E)$ hold, and that the payoff function $h$ fulfills Assumption $(H_{2})$. Then the error term in Theorem \ref{secondorder} satisfies to the following estimate:
\begin{align}\nonumber
|{\rm Error}|\lc \big(\|h^{(1)}(X^M_T)\|_2+&\int_0^1 \|h^{(1)}((1-v)X^M_T+v X_T)\|_3 dv \big)\\ \label{eq:OrderTheoremVanilla}
&\times \frac{M_0}{\sigma_{inf}} (M_1\sqrt T)\big((M_0\sqrt T)^2 +M_J^2\sqrt{\lambda T}\big).
\end{align}
\end{theorem}
The shape of the upper bound regarding $h$ is used for convenience in the proof. In view of the growth condition on $h^{(1)}$, the two first terms depending on $h^{(1)}$ are finite and uniformly bounded as $M_0, M_1, M_J, \lambda$ and$T$ go to 0. 

Analogously to the smooth case (Theorem \ref{OrderTheoremReg}), the approximation error in (\ref{approxfor}) is of order 3 w.r.t. the amplitudes of the model data, meaning that (\ref{approxfor}) is a second order approximation formula.

\begin{proof}
We split the proof into several steps. First, we assume that the payoff is smooth and we establish estimates that depend only on $h^{(1)}$, the first derivative of $h$. For this, we need extra tools from Malliavin calculus, together with tight estimates on the Malliavin derivatives of the parameterized process. Then, we apply a density argument to approximate $h$ under $(H_2)$ by a sequence of smooth payoffs.
\paragraph{Step 1. Malliavin calculus.} For the usual Malliavin calculus on the Wiener space, we refer to Nualart \cite{Nua06}. But our case is slighty different because of jumps. However, in the following, our Malliavin differentiation is w.r.t. the Brownian motion $W$ and not w.r.t. the Poisson measure $\kappa$. Hence formally, it is performed by leaving the jump component fixed, computing the Malliavin derivatives or integration by parts w.r.t. $W$, and then integrating out w.r.t. the jumps. This principle has been formalized in several papers, for instance in \cite{bouc:elie:08} Section 3. We briefly recall a few facts using their notations. \\
The model jumps are associated with the Poisson measure $\kappa$, with intensity $g_{\eta_J,\gamma_J}(x)dx\ \lambda dt$, where $g_{\eta_J,\gamma_J}$ is the Gaussian density on $\R$ with mean $\eta_J$ and variance $\gamma^2_J$. The set of integer-valued measures on $[0,T]\times \R$ is denoted by $\Omega_\kappa$. For $l(.)\in {\mathcal L}=\L_2([0,T],\R)$, the Wiener stochastic integral $\int_0^T l(t)dW_t$ is denoted by $W(l)$. Let ${\cal S}$ denote the class of simple random variables of the form $F=f(W(l_1),\ldots,W(l_N);\kappa)$ where $N\geq1$, $(l_1,\ldots,l_N) \in {\mathcal L}^N$, $f:\R^N\times \Omega_\kappa\mapsto\R$ is bounded and infinitely differentiable w.r.t. its $N$ first components (with bounded derivatives). We denote by $D$ the Malliavin derivative operator with respect to the Brownian motion. For $F \in {\cal S}$, it is defined as the $\mathcal L$-valued random variable given by
$$D_t F=\sum_{i=1}^N \partial_{x_i}
f(W(l_1),\ldots,W(l_N);\kappa)l_i(t).$$ 
The operator $D$ is closable as an
operator from $\L_p(\Omega)$ to $\L_p(\Omega,\mathcal L)$, for any $p\geq 1$. Its domain is denoted by $\D^{1,p}$ with respect to the norm $\|\cdot \|_{1,p}$ given by
$\|F\|^p_{1,p}=\E|F|^p+\E(\int_0^T |D_t F|^2 dt)^{p/2}.$ We can define the iteration of the operator $D$ in such a way that for a smooth random variable $F$, the
derivative $D^kF$ is a random variable with values in ${\mathcal L}^{\otimes k}$.
As in the case $k=1$, the operator $D^k$ is closable from ${\cal S}\subset
\L_p(\Omega)$ into $\L_p(\Omega;{\mathcal L}^{\otimes k})$, $p\geq 1$. Its domain is denoted by $\D^{k,p}$ w.r.t. the norm $\| F\|_{k,p}=[\E|F|^p+\sum_{j=1}^k \E(\|D^j F \|^p_{{\mathcal L}^{\otimes j}})]^{1/p}.$ With this construction, the operator $D$ enjoys the same properties as the usual operator on the Wiener space (see \cite{bouc:elie:08} for more details). This justifies, in the case under study, the application of the usual results established without jumps (in particular the integration by parts formula and the related general $\L_p$ estimates, see the proof of Lemma \ref{NMllemma}).
\paragraph{Step 2. Estimates of Malliavin derivatives.} Under our regularity assumptions $(R_4)$, we know that for any $t\le T$, any $\epsilon\in[0,1]$ and any $p\ge 1$, we have $X^\epsilon_t \in \D^{4,p}$, $X^\epsilon_{1,t} \in \D^{3,p}$, $X^\epsilon_{2,t} \in \D^{2,p}$, $X^\epsilon_{3,t} \in \D^{1,p}$ (see the arguments in \cite{bouc:elie:08}). Actually, we aim at proving the following tight estimates for any $p\ge 2$:
\begin{align}
  \label{eq:x:D1p}
\E|D_rX^{\epsilon}_{t}|^p&\lc |\sigma|^p_\infty,\\
  \label{eq:xm:D1p}
\E|D_rX^{M}_{t}|^p&\lc |\sigma|^p_\infty,\\
 \label{eq:x:D2p}
\E|D^2_{r,s}X^{\epsilon}_{t}|^p&\lc |\sigma|^p_\infty M_1^p,\\
\label{eq:x:D3p}
\E|D^3_{r,s,u}X^{\epsilon}_{t}|^p&\lc |\sigma|^p_\infty M_1^{2p},\\
  \label{eq:x1:D1p}
\E|D_rX^{\epsilon}_{1,t}|^p&\lc M_0^p,\\
  \label{eq:x1:D2p}
\E|D^2_{r,s}X^{\epsilon}_{1,t}|^p&\lc M_0^pM_1^p,\\
\label{eq:x2:D1p}
\E|D_rX^{\epsilon}_{2,t}|^p&\lc  M_1^p\big((M_0\sqrt T)^p + M_J^p \lambda T\big),\\
\label{eq:x2:D2p}
\E|D^2_{r,s}X^{\epsilon}_{2,t}|^p&\lc  M_0^p M_1^p,\\
\label{eq:x3:D1p}
\E|D_rX^{\epsilon}_{3,t}|^p&\lc  M_1^p\big((M_0\sqrt T)^{2p} + M_J^{2p} \lambda T\big),
\end{align}
uniformly in $(r,s,t,u)\in[0,T]^4$ and $\epsilon\in[0,1]$. Here again, the existence of any moment is easy to establish and we will skip the details. We prefer to focus on the dependence of the upper bounds w.r.t. $M_0, M_1, M_J, \lambda$ and $T$. The bounds (\ref{eq:x1:D2p}-\ref{eq:x2:D2p}-\ref{eq:x3:D1p}) are not used for vanilla payoffs, but only for binary ones.\\
{\em Proof of (\ref{eq:x:D1p}).} For $r>t$, $D_rX^{\epsilon}_{t}=0$. Now take $r\le t$, in this case $(D_rX^{\epsilon}_{t})_{r\le t\le T}$ solves the following SDE (see \cite{bouc:elie:08}):
\begin{align}\label{eq:drxe}
  D_rX^{\epsilon}_{t}=\epsilon \sigma_r(X^{\epsilon}_{r-})+\int_r^t D_rX^{\epsilon}_{u-} \epsilon(\sigma^{(1)}_{u}(X^{\epsilon}_{u-})dW_u+\mu^{(1)}_{u}(X^{\epsilon}_{u-})du),
\end{align}
which defines a continuous process. Now, we proceed as in the proof of (\ref{eq:x1:Lp}-\ref{eq:x2:Lp}-\ref{eq:x3:Lp}), combining BDG inequalities and Gronwall's lemma. This gives 
$$\E|D_rX^{\epsilon}_{t}|^p\lc  \E|\sigma_r(X^{\epsilon}_{r-})|^p + T^{p/2-1} M_1^p \int_0^T \E|D_rX^{\epsilon}_{u}|^p du\lc |\sigma|^p_\infty,$$
and proves the announced inequality. Besides, in light of (\ref{eq:xm}) one has $D_rX^{M}_{t}={\mathbf 1}_{r\le t}\sigma_r$, which directly gives (\ref{eq:xm:D1p}).\\
{\em Proof of (\ref{eq:x:D2p}).} Take for instance $r< s\le T$, the other cases are handled in the same way. We have
\begin{align*}
  D^2_{r,s}X^{\epsilon}_{t}=&\epsilon D_r X^{\epsilon}_{s-}\sigma^{(1)}_s(X^{\epsilon}_{s-})\\
&+\int_s^t D_{s}X^{\epsilon}_{u-} \epsilon(D_r X^{\epsilon}_{u-}\sigma^{(2)}_{u}(X^{\epsilon}_{u-})dW_u+D_r X^{\epsilon}_{u-}\mu^{(2)}_{u}(X^{\epsilon}_{u-})du)\\
&+\int_s^t D^2_{r,s}X^{\epsilon}_{u-} \epsilon(\sigma^{(1)}_{u}(X^{\epsilon}_{u-})dW_u+\mu^{(1)}_{u}(X^{\epsilon}_{u-})du),
\end{align*}
which implies, in particular, that $t\mapsto D^2_{r,s}X^{\epsilon}_{t}$ is continuous. It readily follows that 
\begin{align*}
\E|D^2_{r,s}X^{\epsilon}_{t}|^p&\lc  M_1^{p}\E|D_r X^{\epsilon}_{s}|^p +T^{p/2} M_1^{p} \sup_{r<s\le u\le T}\E|D_{s}X^{\epsilon}_{u}D_{r}X^{\epsilon}_{u}|^{p}\\
&\lc M_1^p \E|D_r X^{\epsilon}_{s}|^p +T^{p/2} M_1^{p} \sup_{r<s\le u\le T}(\E|D_{s}X^{\epsilon}_{u}|^{2p}+\E|D_{r}X^{\epsilon}_{u}|^{2p})\lc |\sigma|^p_\infty M_1^p
\end{align*}
where we have used the  Young inequality $ab\le \frac 12( a^2+b^2)$ in the second line and (\ref{eq:x:D1p}) in the last inequality. The estimate (\ref{eq:x:D3p}) can be established in the same way.\\
{\em Proof of (\ref{eq:x1:D1p}).} We only consider $r\le t$. Here one has
\begin{align*}
  D_rX^{\epsilon}_{1,t}=&\sigma_r(X^{\epsilon}_{r-})+\epsilon X^{\epsilon}_{1,r-}\sigma^{(1)}_{r}(X^{\epsilon}_{r-})\\
&+\int_r^t D_rX^{\epsilon}_{u-} (\sigma^{(1)}_{u}(X^{\epsilon}_{u-})+\epsilon  X^{\epsilon}_{1,u-}\sigma^{(2)}_{u}(X^{\epsilon}_{u-}))dW_u\\
&+\int_r^t D_rX^{\epsilon}_{u-} (\mu^{(1)}_{u}(X^{\epsilon}_{u-})+\epsilon  X^{\epsilon}_{1,u-}\mu^{(2)}_{u}(X^{\epsilon}_{u-}))du\\
&+\int_r^t D_rX^{\epsilon}_{1,u-} \epsilon(\sigma^{(1)}_{u}(X^{\epsilon}_{u-}) d{W}_{u}+\mu^{(1)}_{u}(X^{\epsilon}_{u-}) du).
\end{align*}
It readily follows that 
\begin{align*}
  \E|D_rX^{\epsilon}_{1,t}|^p\lc \E|\sigma_r(X^{\epsilon}_{r-})+\epsilon X^{\epsilon}_{1,r-}\sigma^{(1)}_{r}(X^{\epsilon}_{r-})|^p
+T^{p/2} M_1^p \sup_{r\le u\le T} (\E|D_rX^{\epsilon}_{u}|^p+\E|D_rX^{\epsilon}_{u}X^{\epsilon}_{1,u}|^p).
\end{align*}
Since a fixed time $r$ is equal to a jump time with null probability and thanks to the Young inequality, we obtain
\begin{align*}
  \E|D_rX^{\epsilon}_{1,t}|^p\lc |\sigma|^p_\infty+M_1^p\E|X^{\epsilon}_{1,r}|^p+T^{p/2} M_1^p \sup_{r\le u\le T} (\E|D_rX^{\epsilon}_{u}|^p+\E|D_rX^{\epsilon}_{u}|^{2p}+ \E|X^{\epsilon}_{1,u}|^{2p}).
\end{align*}
It remains to take advantage of the inequalities (\ref{eq:x1:Lp}) and (\ref{eq:x:D1p}), and to use $|\sigma|_\infty\le M_0$ and $M_1\le M_0$ to complete the proof of (\ref{eq:x1:D1p}).\\
{\em Proof of (\ref{eq:x1:D2p}-\ref{eq:x2:D1p}-\ref{eq:x2:D2p}-\ref{eq:x3:D1p}).} They can be proved similarly, with long and tedious computations. Since there is no extra difficulty, we will skip further details.
\paragraph{Step 3. Bounding the error using only $h^{(1)}$, when $h$ is smooth.}
We come back to the representation (\ref{eq:error:smooth}) for the error. The first term can be estimated using a Cauchy-Schwartz inequality and (\ref{eq:x3:Lp}):
\begin{align}
\nonumber|\E[h^{(1)}(X^M_T)\overline X_{3,T}]|&\lc \|h^{(1)}(X^M_T)\|_2 \sup_{\epsilon\in[0,1]} \|X_{3,T}^\epsilon\|_2\\
&\lc \|h^{(1)}(X^M_T)\|_2
(M_1\sqrt T)\big((M_0\sqrt T)^{2} + M_J^{2} \sqrt{\lambda T}\big).\nonumber
%\label{eq:term1:vanilla}
\end{align}
This fits the required upper bound (\ref{eq:OrderTheoremVanilla}) well, because $M_0\ge \sigma_{inf}$.

The second term in (\ref{eq:error:smooth}) requires a little extra work because of $h^{(2)}$. For this, we state a lemma, proof of which is given at the end.
\begin{lemma}\label{NMllemma}
Assume (E) and ($R_{3}$). Let $Z$ belong to $\cap_{p\ge 1}\D^{2,p}$. For any $v \in [0,1]$, for $k=1,2$, there exists a random variable $Z^{v}_{k}$ in any $L_p$ ($p\ge 1$) such that for any function $l\in \C^{\infty}_{0}(\R)$, one has  
$$\E[ l^{(k)}(v X_{T}+(1-v)  X^M_T)Z]=\E[ l(v X_T+(1-v)  X^M_T)Z^{v}_{k}].$$
Moreover, one has $\|Z^{v}_{k}\|_{p}\lc \frac{\|Z\|_{k,p+\frac{1}{2}}}{(\sigma_{inf} \sqrt{T})^{k}},$ uniformly in $v$.
\end{lemma}
Apply this Lemma with $k=1$ and $Z=(\overline X_{2,T})^2$ defined in (\ref{eq:bar:x}). From the estimates (\ref{eq:x2:Lp}-\ref{eq:x2:D1p}), we readily obtain $$\|Z^{v}_{1}\|_{\frac{3}{2}}\lc \frac{(M_1\sqrt T)^2\big((M_0\sqrt T)^2 + M^2_J \sqrt{\lambda T}\big)}{\sigma_{inf} \sqrt{T}}.$$
We have proved the upper bound (\ref{eq:OrderTheoremVanilla}).
\paragraph{Step 4. Bounding the error under the sole assumption $(H_2)$.}
So far, our error estimates depend on $h^{(1)}$, but they have been established for smooth payoffs $h$. It remains to justify that the error upper bound still holds for payoffs that are only almost everywhere differentiable (assumption $(H_2)$). We argue by regularization, which is somewhat standard but a bit tricky here. We follow the proof of \cite{GM05}.  

Denote by $\rho$ the measure defined by $\int_{\R}g(x)\rho(dx)=\E(g( X_{T}))+\E(g(X^M_T)) +\E(g(X^M_T+Y')) +\int_{0}^{1}\E(g(v X_{T} +(1-v) X^M_T))dv$. It is well known (see \cite{Rud66} for instance) that there exists a sequence $(h_{n})_{n \in \N}$ of smooth  functions converging to $h$ in $\L_{3}(\rho)$ as well as its first derivative, as $n$ goes to infinity. Thus, we can pass to the limit for $\E(h_n(X_T))$ and $\E(h_n(X^M_T))$. In view of (\ref{eq:OrderTheoremVanilla}), we can also pass to the limit for the error bound. It remains to pass to the limit for the corrections terms, i.e. for the greeks $\G^{h_n}_{i}(X^M_T)$ and $\G^{h_n}_{i}(X^M_T+Y')$. To accomplish this, we represent them as $\E(h_n(X^M_T)Z_i)$ and $\E(h_n(X^M_T+Y')Z_i)$ using Lemma \ref{NMllemma} with $Z=1$. Since $Z_i$ is in $\L_{3/2}$, we can pass to the limit as $n\rightarrow \infty$ to get $\E(h(X^M_T)Z_i)=\G^{h}_{i}(X^M_T)$ and $\E(h(X^M_T+Y')Z_i)=\G^{h}_{i}(X^M_T+Y')$.\qed
\end{proof}
%We mention that the exponent $1/4$ of $\lambda T$ in (\ref{eq:OrderTheoremVanilla}) can be improved to $1/2$, by applying H\"older inequalities in Step 4 of the above Theorem and considering better exponents for H\"older inequalities in Step 2 of the proof of Lemma \ref{NMllemma}.
\paragraph{Proof of Lemma \ref{NMllemma}.}
Take $k=1$ or $2$.
\paragraph{Step 1. $F_{v}=v X_{T}+(1-v) X^M_T$ is a non degenerate random variable (in the Malliavin sense).} Under $(R_4)$, we know that $F_{v}$ is in $\cap_{p\ge 1}\D^{4,p}$. One has to prove that $\gamma _{F_v}=\int_{0}^{T}(D_{s}F_v)^{2}ds$ is almost surely positive and its inverse is in any $\L_p$ ($p\ge 1$). From the linear SDE (\ref{eq:drxe}) satisfied by $(D_s X_t)_{s\le t\le T}$, we obtain
\begin{align*}
\gamma _{F_v}&=\int_{0}^{T}(v \sigma_s(X_{s-})e^{\int_{s}^{T}\sigma^{(1)}_u(X_{u-})dW_{u}+(\mu^{(1)}_u-\frac 12(\sigma^{(1)}_u)^2)(X_{u-})du} +(1-v)\sigma_s(x_{0}) )^{2}ds,
\end{align*}
which clearly leads to our claim. Besides, for any $p\geq 1$, we derive
\begin{align*}
\|\gamma^{-1}_{F_v}\|_{p}\lc ( \sigma_{inf} \sqrt T)^{-2}.
\end{align*}
\paragraph{Step 2. Integration by Parts formula.} Using Proposition 2.1.4 and Proposition 1.5.6 in \cite{Nua06}, one gets the existence of $Z^{v}_{k}$ in $\L_{p}$ with
\begin{align*}
\|Z^{v}_{k}\|_{p}\lc \|\gamma^{-1}_{F_v}\|^{k}_{k,2^{k}p(2p+1)}\|DF_v\|^{k}_{k,2^{k}p(2p+1)}\|Z\|_{k,p+\frac{1}{2}}.
\end{align*} 
\paragraph{Step 3: Upper bound of $\|DF_v\|_{k,q},\|\gamma^{-1}_{F_v}\|_{k,q}$ for $q\geq 2$. }
On the one hand, using the inequalities (\ref{eq:x:D1p}-\ref{eq:xm:D1p}-\ref{eq:x:D2p}-\ref{eq:x:D3p}), we easily obtain 
\begin{equation}
  \label{eq:dfv}
\|DF_{v}\|_{k,q}\lc |\sigma|_{\infty} \sqrt{T}.
\end{equation}
On the other hand, with the same inequalities, we get $\sup_{r\le T}\E|D_{r} \gamma_{F_v}|^p\lc T^p|\sigma|_\infty^{2p}M_1^{p}$ and $\sup_{r,s\le T}\E|D^2_{r,s} \gamma_{F_v}|^p \lc T^p|\sigma|_\infty^{2p}M_1^{2p}$ for any $p\ge 2$. Then, after some computations, it follows that
\begin{equation}
  \label{eq:gamma:inverse}
\|\gamma^{-1}_{F_v}\|_{2,p}\lc (\sigma_{inf} \sqrt T)^{-2}\big(1+\frac{M_1|\sigma|_\infty^{2}T^{1/2}}{\sigma^{2}_{inf}}+\frac{M_1^{2}|\sigma|_\infty^{2}T}{\sigma^{2}_{inf}}+\frac{M_1^{2}|\sigma|_\infty^{4}T}{\sigma^{4}_{inf}}\big)
\end{equation}
for any $p\ge 2$. Finally using $|\sigma|_\infty\le C_E \sigma_{inf}$ (assumption $(E)$) combined with (\ref{eq:dfv}) and (\ref{eq:gamma:inverse}), we get 
\begin{equation*}\|\gamma^{-1}_{F_v}\|^{k}_{k,2^{k}p(2p+1)}\|DF_v\|^{k}_{k,2^{k}p(2p+1)}\lc (\sigma_{inf}\sqrt{T})^{-2k} (|\sigma|_{\infty} \sqrt{T})^k  \lc (\sigma_{inf}\sqrt{T})^{-k}.\end{equation*}
This completes our proof. \qed 

\subsection{Error analysis for binary payoff (under ($H_3$))}\label{subBO}
For this kind of option, the payoff $h$ is not necessarily smooth. We only assume that $h$ is in $\Hh$. The results below are easy extensions of the vanilla options case, we leave the proof to the reader.
\begin{theorem}\label{OrderTheoremReg0} \textbf{Error for binary payoff.} Assume that $(R_{4})$ and $(E)$ hold, and that the payoff function $h$ fulfills Assumption $(H_{3})$. Then the error term in Theorem \ref{secondorder} satisfies the following estimate:
\begin{align*}
|{\rm Error}|\lc \big(\|h(X^M_T)\|_3+&\int_0^1 \|h((1-v)X^M_T+v X_T)\|_3 dv \big)\\ 
&\times \big(\frac{M_1}{\sigma_{inf}}+\frac{M^2_1}{\sigma^2_{inf}}\big) \big((M_0\sqrt T)^2 +M_J^2\sqrt{\lambda T}\big).
\end{align*}
\end{theorem}
Unlike the cases of smooth and vanilla payoff, for binary payoffs the approximation formula (\ref{approxfor}) is of first order w.r.t. the amplitudes of the model data (with error terms of order 2). This is inherent to the lack of regularity of the payoff.

\section{Appendix}\label{sectApp}
\subsection{Technical results related to explicit correction terms}
In this subsection, we bring together the results (and their proofs) which allow us to derive the explicit terms in the formula (\ref{approxfor}).\\
In the following, $(u_{t})$ (resp. $(v_{t})$ and $(\nu_{t})$) are  square integrable and predictable (resp. deterministic) process and $l$ is a smooth function with compact support.\\
\begin{lemma}\label{lemma1}
For any continuous (or piecewise continuous) function $f$, any continuous semimartingale $Z$ vanishing at t=0, one has:
\begin{equation*}
\int_{0}^{T}f_{t}Z_{t}dt=\int_{0}^{T}(\int_{t}^{T}f_{s}ds)dZ_{t}.
\end{equation*}
\end{lemma}
\begin{proof} This follows from the It\^ o formula applied to the product $(\int_{t}^{T}f_{s}ds)Z_{t}$.
\qed \end{proof}
\begin{lemma}\label{lemma2}
One has:
\begin{equation*}
\E[(\int_{0}^{T}u_{t}dW_{t})l(\int_{0}^{T}v_{t}d{W}_{t})]=\E[(\int_{0}^{T}v_{t}u_{t}dt)l^{(1)}(\int_{0}^{T}v_{t}d{W}_{t})].
\end{equation*}
In the case of deterministic $u$, it is equal to $\int_{0}^{T}v_{t}u_{t}dt\hspace{0.05cm}\G^{l}_{1}(\int_{0}^{T}v_{t}d{W}_{t})$.
\end{lemma}
\begin{proof}
We first give the proof in a particular case when $u$ and $v$ are equal to 1. By a usual integration by parts formula, one has:
\begin{align*}
\E[l(W_{T})W_{T}]=\int_{-\infty}^{\infty} l (\sqrt{T}x)\sqrt{T}x \frac{e^{\frac{-x^{2}}{2}}}{\sqrt{2\pi}}dx
=\int_{-\infty}^{\infty}  T l^{(1)} (\sqrt{T}x) \frac{e^{\frac{-x^{2}}{2}}}{\sqrt{2\pi}}dx
=T\E[l^{(1)}(W_{T})].
\end{align*}
For the general proof: apply the duality relationship of Malliavin calculus (see Lemma 1.2.1 in \cite{Nua06}), identifying It\^o's integral and Skorohod operator for adapted integrands.\qed
\end{proof}
\begin{lemma}\label{corollary2} Write $(X^{c}_{1,t})_t$ for the continuous part of $(X_{1,t})_t$. One has :
\begin{align*} 
\E[(\int_{0}^{T}\nu_{t}X^{c}_{1,t}dt)l(\int_{0}^{T}v_{t}d{W}_{t})]
&=\int_{0}^{T} v_{t} \sigma_{t} (\int_{t}^{T}\nu_{s}ds) dt\hspace{0.05cm} \G^{l}_{1}(\int_{0}^{T}v_{t}d{W}_{t})\\&+\int_{0}^{T}  \mu_{t} (\int_{t}^{T}\nu_{s}ds) dt\hspace{0.05cm} \G^{l}_{0}(\int_{0}^{T}v_{t}d{W}_{t}).
\end{align*}
\end{lemma}
\begin{proof}
Applying first Lemma \ref{lemma1} to $f(t)=\nu_{t}$ and $Z_{t}=X^c_{1,t}$, one has: 
\begin{align*}
\E[(\int_{0}^{T}\nu_{t}X^{c}_{1,t}dt)l(\int_{0}^{T}v_{t}d{W}_{t})]&=\E[(\int_{0}^{T} (\int_{t}^{T}\nu_{s}ds)dX^{c}_{1,t}) l(\int_{0}^{T}v_{t}d{W}_{t})]\\
&=\E[(\int_{0}^{T}  (\int_{t}^{T}\nu_{s}ds)(\sigma_{t} d{W}_{t}+\mu_{t} dt) l(\int_{0}^{T}v_{t}d{W}_{t})]\\
&=(\int_{0}^{T} v_{t} \sigma_{t} (\int_{t}^{T}\nu_{s}ds) dt) \E[l^{(1)}(\int_{0}^{T}v_{t}d{W}_{t})]\\&+(\int_{0}^{T}  \mu_{t}  (\int_{t}^{T}\nu_{s}ds)dt) \E[l(\int_{0}^{T}v_{t}d{W}_{t})],
\end{align*}
and we have used Lemma \ref{lemma2} for the last equality. \qed
\end{proof}
\begin{lemma}\label{LJKcalculus}
One has:
\begin{align*} 
\E[(\int_{0}^{T}\nu_{t}J_{t}dt)l(J_{T})]&=\lambda(\eta_{J}\int_{0}^{T}t \nu_{t}dt\hspace{0.05cm}\G^{l}_{0}(J_{T}+Y')\\&+\gamma_{J}\int_{0}^{T}t \nu_{t}dt\hspace{0.05cm}\G^{l}_{1}(J_{T}+Y')),
\end{align*}
such that $Y'$ is an independent copy of the variables $(Y_{i})_{i\in \N^*}$.
\end{lemma}
\begin{proof}
Using the independence of increments for $J$, one has:
\begin{align*} 
\E[(\int_{0}^{T}\nu_{t}J_{t}dt)l(J_{T})]=\int_{0}^{T}\nu_{t}\E[J_{t}l(J_{T}-J_{t}+J_{t})]dt=\int_{0}^{T}\nu_{t}\E[\iota(J_{T}-J_{t})]dt.
\end{align*}
Using a conditioning argument and since $\sum_{j=1}^{k}Y_{j}$ is a Gaussian random variable, one has:
\begin{align*}
\iota(x)=\E[J_{t}l(x+J_{t})]&
        =\sum_{k \in \N^{*}}\P(N_{t}=k)\E[\sum_{j=1}^{k}Y_{j}l(x+\sum_{j=1}^{k}Y_{j})]\\&
        =\sum_{k \in \N^{*}}\P(N_{t}=k)k(\eta_{J}\E[l(x+\sum_{j=1}^{k}Y_{j})]+\gamma_{J}\E[l^{(1)}(x+\sum_{j=1}^{k}Y_{j})])  \\&
        =\sum_{k \in \N}\lambda t\P(N_{t}=k)  (\eta_{J}\E[l(x+\sum_{j=1}^{k+1}Y_{j})]+\gamma_{J}\E[l^{(1)}(x+\sum_{j=1}^{k+1}Y_{j})])\\&
        =\lambda t(\eta_{J}\E[l(x+J_{t}+Y')]+\gamma_{J}\E[l^{(1)}(x+J_{t}+Y')]),
\end{align*}
with $Y'$ as in the lemma statement.\qed
\end{proof}
% \subsection{Fa\`a di Bruno's formula}
% \begin{lemma}\label{BrunoLemma}
% If g and f are functions with a sufficient number of derivatives, then
% \begin{align*}
% (g(f(\epsilon)))^{(n)}=\mathop{\sum_{k=(k_{1},\cdots,k_{n})\in \N^{n}}}_{\sum_{j=1}^{n}jk_{j}=n}d_{k}g^{(\sum_{j=1}^{n}k_{j})}(f(\epsilon))\prod_{j=1}^{n} (f^{(j)}(\epsilon))^{k_{j}},
% \end{align*}
% where $d_{k}$ are integer numbers depending only on $k$. Notice that when $k_{n}=1$ one has $d_{k}=1$.
% \end{lemma}
% The proof of this Lemma can be found in \cite{Fadi1857}.
\subsection{Upper bound for compound Poisson process}
\begin{lemma}\label{Jlem}
The $\L_{p}$ norm ($p\ge 1$) of the compound Poisson process at time $t\le T$ can be estimated as follows:
\begin{align*}
\E|J_t|^p \lc M_J^p \lambda t.
\end{align*}
\end{lemma}
\begin{proof} Set $Z_j=(Y_j-\eta_J)/\gamma_J$. The random variables $(Z_j)_j$ are i.i.d. Gaussian variables, with zero mean and unit variance. Then
\begin{align*}
|J_{t}|&=|\sum_{j=1}^{N_{t}} \eta_J+\gamma_J Z_j|
 \leq |\eta_J|N_t+\gamma_J|\sum_{j=1}^{N_{t}} Z_j|\le M_J (N_t+|\sum_{j=1}^{N_{t}} Z_j|).
\end{align*}
Now it only remains to compute the $p$-th moment of $N_t$ and $K_t=|\sum_{j=1}^{N_{t}} Z_j|$, which is considered a standard exercise. We give few details about the second term $K_t$. First compute the characteristic function $\varphi(u)=\E(e^{iu \sum_{j=1}^{N_{t}} Z_j})=\exp(\lambda t (e^{-u^2/2}-1))$. Then for an even integer $p$, one has $\E(\sum_{j=1}^{N_{t}}Z_j)^p=\E(K_t^p)=i^p  \varphi^{(p)}(0)=O(\lambda t)$. For odd values of $p$ of the form $p=2k+1$, we apply the inequality $ab\le \frac 12( a^2+b^2)$ to write $K_t^p\le \frac 12 (K_t^{2k}+K_t^{2k+2})$. The result then follows by using the estimates from the previous case ($p$ even).\qed
\end{proof}
\bibliography{smartexpansionjump}
\bibliographystyle{alpha}
\end{document}